\documentclass[a4paper,11pt]{article}
\usepackage{pos}

\title{Thermal lattice QCD results from the FASTSUM collaboration}

\author*[a]{Chris Allton}
\author[a]{Gert Aarts}
\author[a]{M.~Naeem Anwar}
\author[a,f]{Ryan Bignell}
\author[a]{Timothy J.~Burns}
\author[a]{Sergio Chaves Garc\'ia-Mascaraque}
\author[b]{Simon Hands}
\author[c]{Benjamin J\"ager}
\author[d]{Seyong Kim}
\author[e]{Maria Paola Lombardo}
\author[a]{Benjamin Page}
\author[f]{Sin\'ead Ryan}
\author[g]{Jon-Ivar Skullerud}\author[a]{Antonio Smecca}
\author[a]{Thomas Spriggs}

\affiliation[a]{Department of Physics, Swansea University, Swansea, SA2 8PP, United Kingdom}
\affiliation[b]{Department of Mathematical Sciences, University of Liverpool, Liverpool L69 3BX, United Kingdom}
\affiliation[c]{CP3-Origins \& Danish IAS, Department of Mathematics and Computer Science, University of Southern Denmark, 5230, \
Odense M, Denmark}
\affiliation[d]{Department of Physics, Sejong University, Seoul 05006, Korea}
\affiliation[e]{INFN, Sezione di Firenze, 50019 Sesto Fiorentino (FI), Italy}
\affiliation[f]{School of Mathematics and Hamilton Mathematics Institute, Trinity College Dublin, Ireland}
\affiliation[g]{Department of Theoretical Physics, National University of Ireland Maynooth, County Kildare, Ireland}

\emailAdd{c.r.allton@swansea.ac.uk}

\abstract{The FASTSUM Collaboration has developed a comprehensive research programme in thermal lattice QCD using 2+1 flavour ensembles. We review our recent hadron spectrum analyses of open charm mesons and charm baryons at non-zero temperature. We also detail our determination of the interquark potential in the bottomonium system using NRQCD quarks.
All of our work uses anisotropic lattices where the temporal lattice spacing is considerably finer than the spatial one allowing better resolution of temporal correlation functions.}

\FullConference{42nd International Conference on High Energy Physics (ICHEP2024)\\
18-24 July 2024\\
Prague, Czech Republic\\}

\begin{document}
\maketitle

\section{Introduction}

The QCD phase diagram is an active area of research which has wide applications to our understanding of the early universe, dense astronomical objects and heavy-ion collision experiments.
The {\sc fastsum} collaboration has been actively studying thermal QCD using the lattice approach for many years.
We have successfully calculated many quantities at zero chemical potential, $\mu = 0$
\cite{
Aarts:2022krz,
Aarts:2023nax,
Aarts:2015mma,
Aarts:2017rrl,
Aarts:2018glk,
Aarts:2020vyb,
Aarts:2014cda,
Aarts:2014nba}
and have made some tentative steps into the interior of the phase diagram, i.e. at $\mu\ne 0$ \cite{Nikolaev:2020vll}.

In these proceedings, we present a summary of some of our collaboration's thermal QCD results at $\mu=0$.
We begin with a brief outline of our approach using anisotropic lattices. Subsequent sections summarise our results for open charm mesons, charm baryons and the interquark potential in bottomonium.


\section{{\sc fastsum} approach}

The {\sc fastsum} collaboration has an established programme of obtaining spectral features and other physical properties of QCD at non-zero temperature from lattice simulations.
Our specific approach uses {\em anisotropic} lattices, where the temporal lattice spacing is much finer than the spatial one, i.e. $a_\tau \ll a_s$. This enables us to extract more information from time-dependent quantities such as temporal correlation functions which are fundamental to the spectral analyses considered here. Table~\ref{tb:latt_params} contains a summary of the parameters of our simulations including the temperatures.

\begin{table}[h]
  \centering
    \caption{The \textsc{FASTSUM} Generation 2L ensembles used in this work.
    The spatial lattice spacing is $a_s = 0.11208 \left(31\right)$ fm,
    with a renormalised anisotropy $\xi = a_s/a_\tau = 3.453(6)$.
    Lattice volumes are $32^3 \times N_\tau$ and
    temperature $T = 1/\left(a_\tau N_\tau\right)$.
    Our pion mass is $239(1)$ MeV~\cite{Wilson:2019wfr}.
    We have $\sim 1000$ configurations at each $T$.
    The pseudocritical point is $T_{pc} = 167(3)$, indicated by the double vertical lines.
    Full details of these ensembles may be found in \cite{Aarts:2020vyb,Aarts:2022krz}.}
  \begin{tabular}{r|rrrrr||rrrrrr}
  $N_\tau$ & 128 & 64 & 56 & 48 & 40 & 36 & 32 & 28 & 24 & 20 & 16\\ \hline
  $T\, \left(\text{MeV}\right)$ & 47 & 95 & 109 & 127 & 152 & 169 & 190 & 217 & 253 & 304 & 380 \\
  $T/T_{pc}$ &\!\!\! 0.284 &\!\!\! 0.569 &\!\!\! 0.650 &\!\!\! 0.758 &\!\!\! 0.910 &\!\!\!
  1.011 &\!\!\! 1.138 &\!\!\! 1.300 &\!\!\! 1.517 &\!\!\! 1.820 &\!\!\! 2.314
  \end{tabular}
\label{tb:latt_params}
\end{table}


\section{Open Charm Mesons}

In this section we study the open charm meson spectrum in a thermal medium.
At very low temperature where the thermal widths are negligible and the channel is dominated by the ground state, mesonic correlation functions have the ``model'' behaviour,
\[
G_\text{model}(\tau;T,T_0) = Z \frac{\cosh[M(T_0)(\tau-1/2T)]}{\sinh[M(T_0)/2T)},
\]
where $M(T_0)$ is the ground state mass obtained from fits of the correlation data at a reference temperature, $T_0$, and $Z$ is the amplitude. We set $T_0$ to be the lowest temperature studied, i.e. $T_0=47$~MeV.
If $G_\text{model}$ reproduces the data for a range of $T$, then this implies that the ground state mass is temperature independent for this range.

\begin{figure}[t!]
  \centering
  \includegraphics[page=1,width=0.48\columnwidth,keepaspectratio,origin=c]
  {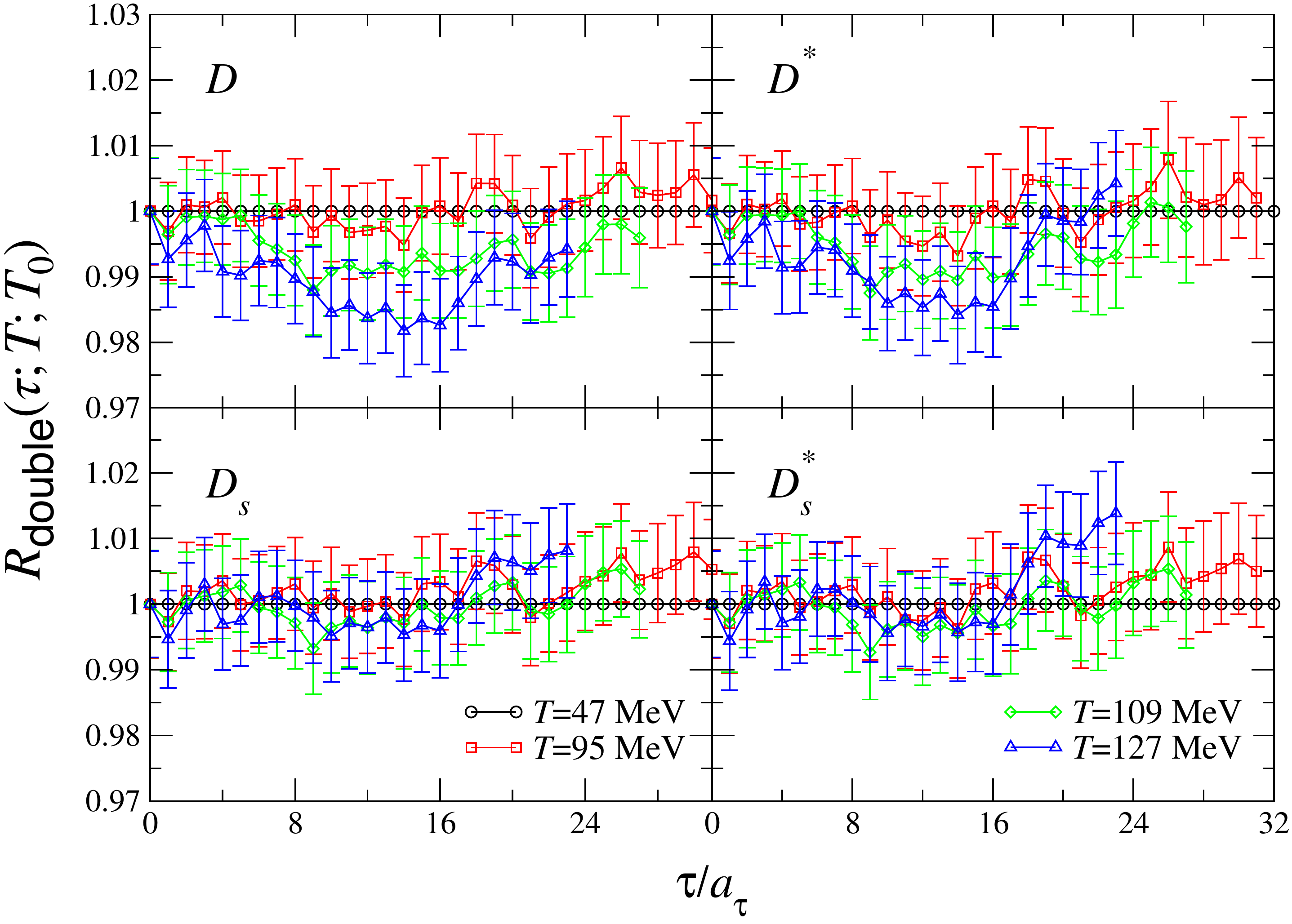}
  \includegraphics[page=1,width=0.48\columnwidth,keepaspectratio,origin=c]
  {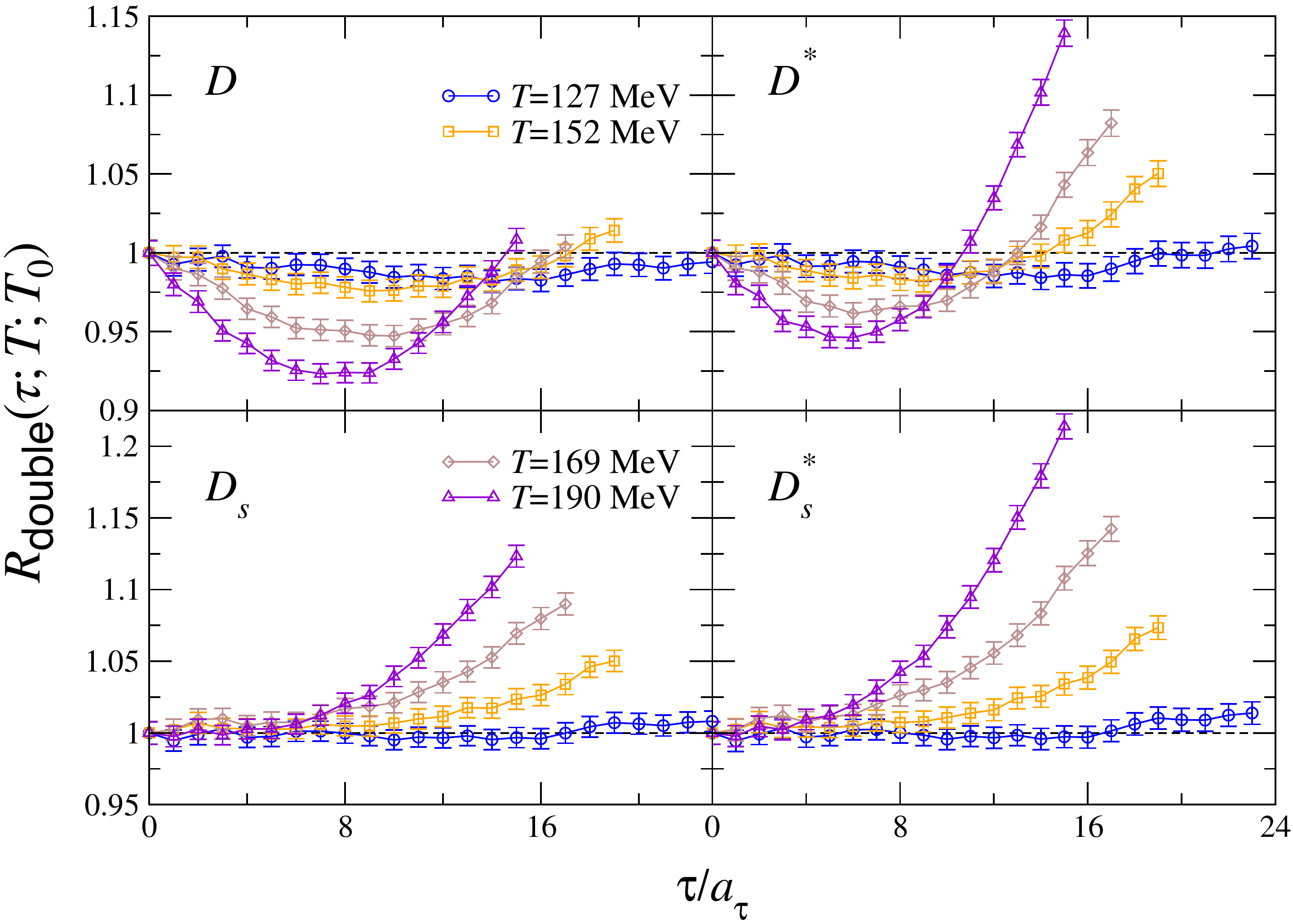}
  \caption{Double Ratio plots for the $D-$meson family at low (left) and intermediate (right) temperatures,
  clearly showing a thermal effect for $T \gtrsim 127$~MeV, see Eq.(\ref{eq:double-ratio}). The reference temperature is set to $T_0 = 47$~MeV.
  }
  \label{fig:double-ratio}
\end{figure}

To quantify this, we study the double ratio,
\begin{equation}
R_\text{double}(\tau;T,T_0) =
\frac{R_\text{model}(\tau;T  ,T_0)}
     {R_\text{model}(\tau;T_0,T_0)}
\;\;\;\;\;\;
\text{where}
\;\;\;\;\;\;
R_\text{model}(\tau;T,T_0) = \frac{G(\tau;T)}{G_\text{model}(\tau;T,T_0)}.
\label{eq:double-ratio}
\end{equation}
$G(\tau;T)$ is the correlation function data obtained from the lattice simulation.
This double ratio is a means of partially cancelling some systematic effects coming from the excited states.
Deviations of $R_\text{double}$ from unity therefore measure thermal variations in the ground state mass.
In Fig.\ref{fig:double-ratio}, we plot $R_\text{double}$ for a variety of open charm mesons.
As can be seen, for temperatures, $T\le 127$~MeV, there is no indication of thermal effects. For intermediate temperatures, $127 \text{ MeV} \le T \le 190 \text{ MeV}$, there are signs of a deviation of $R_\text{double}$ from unity, implying that the ground state mass differs from $M(T_0)$ for these temperatures.

\begin{figure}[t!]
  \centering
  \includegraphics[page=1,width=0.48\columnwidth,keepaspectratio,origin=c]
  {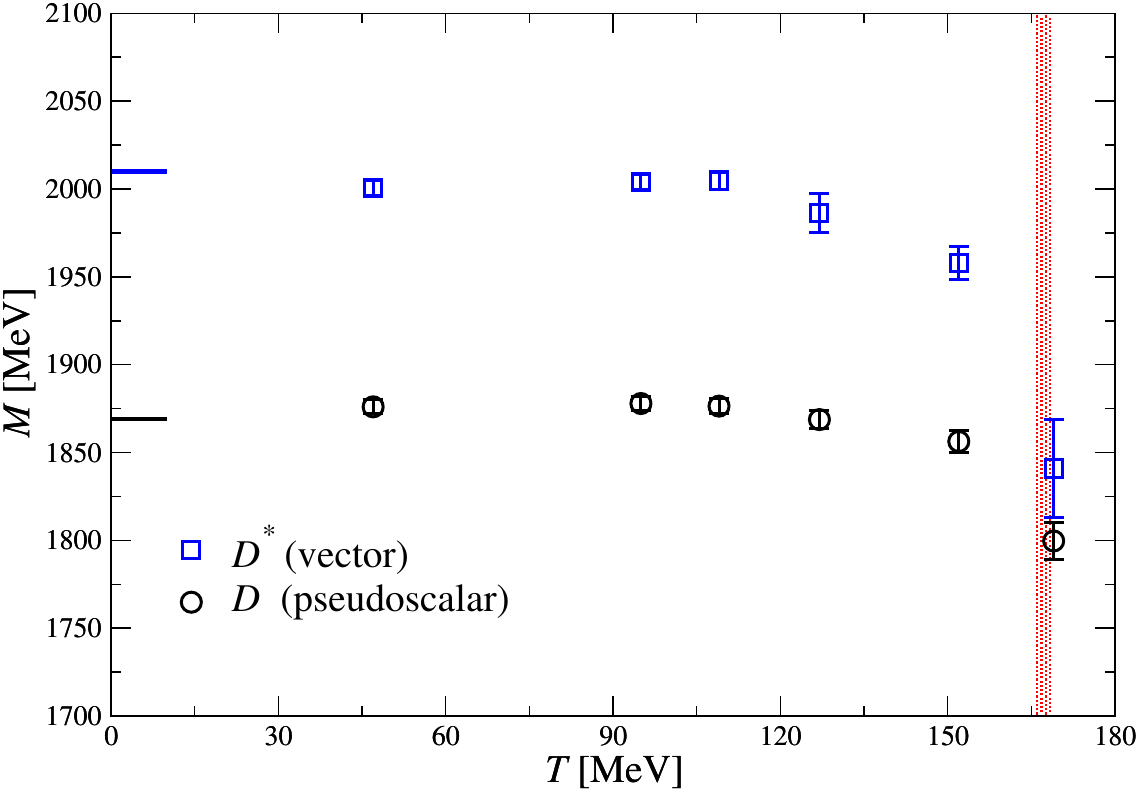}
  \includegraphics[page=1,width=0.48\columnwidth,keepaspectratio,origin=c]
  {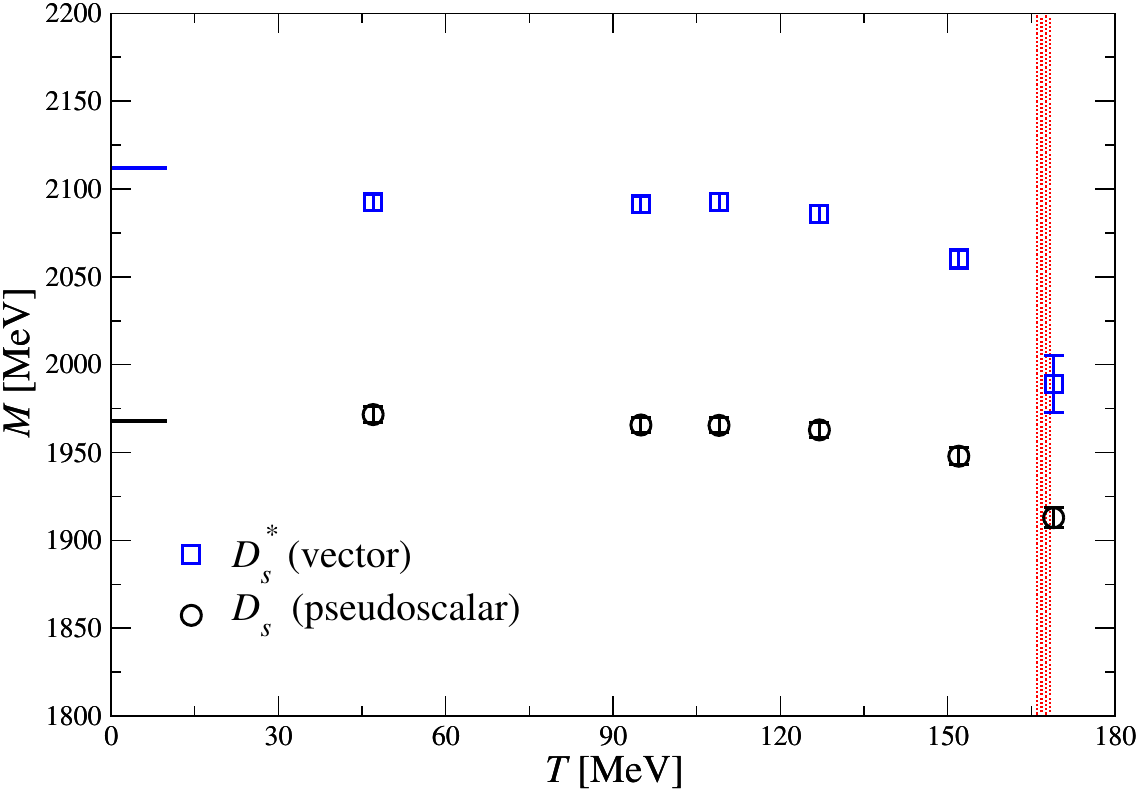}
  \caption{The ground state mass of the $D-$meson family for $T \lesssim T_{pc}$ showing clear thermal variations.
  $T_{pc}=167(3)$~MeV is shown by the vertical band.
  }
  \label{fig:charm-meson-mass-vs-T}
\end{figure}

We also perform fits of the correlation functions to $G_\text{model}(\tau;T,T)$ to obtain the temperature dependent mass $M(T)$ displayed in Fig.\ref{fig:charm-meson-mass-vs-T}.
We restrict this analysis to the temperature range $T\lesssim T_{pc}$ because we do not have confidence in the applicability of $G_\text{model}(\tau;T,T)$ for larger temperatures.

These results are discussed in more detail in \cite{Aarts:2022krz}.


\section{Charm Baryons}

At zero temperature, the parity partners of light baryons in Nature are clearly non-degenerate. E.g. $M_+ = 0.939$ GeV for the nucleon and $M_- = 1.535$ GeV for its negative parity partner.
In situations where there is exact chiral symmetry, it can be shown that these parity partners are necessarily degenerate.
This suggests it would be instructive to study the masses of parity partners above $T_{pc}$ where chiral symmetry is restored.

We note that baryonic correlation functions at temperature, $T=1/(a_\tau N_\tau)$, can be written
\[
G(\tau;N_\tau) = \int_{-\infty}^\infty \frac{d\omega}{2\pi}\, K_F(\tau,\omega;N_\tau)\,\rho(\omega; N_\tau),
\;\;\;\;\;\;
\text{where}
\;\;\;\;\;\;
K_F(\tau,\omega; N_\tau)
= \frac{e^{-\omega a_\tau N_\tau}}{1 + e^{-\omega a_\tau N_\tau}},
\]
where $\rho$ is the spectral function and the fermionic kernel is $K_F$.
We use the approach introduced in \cite{Ding:2012sp}, 
which expresses the kernel at $N_\tau$ as a sum over kernels at other temperatures,
\[
K_F(\tau,\omega; N_\tau)
= \sum_{n=0}^{m-1} (-1)^n \, K_F(\tau + nN_\tau, \omega; mN_\tau),
\]
and defines the ``reconstructed'' correlator,
\[
G_\text{rec}(\tau;\, N_\tau,\, N_0)
= \int_{-\infty}^\infty \frac{d\omega}{2\pi}\, K_F(\tau,\omega;N_\tau)\,\rho(\omega; N_0)
= \sum_{n=0}^{m-1} (-1)^n \, G(\tau + nN_\tau;\, N_0).
\]
Thus $G_\text{rec}(\tau;N_\tau, N_0)$ is comprised of the spectral function at the reference temperature $N_0$, with temperature effects entering only in the ``geometric'' factors in the kernel, $K_F$.
This is analogous to the previous section where we defined $G_\text{model}$ assuming that the ground state mass was unchanged from its value at the reference temperature, $M(T_0)$.
The ratio $G_\text{rec}(\tau; N_\tau, N_0) / G(\tau; N_\tau)$
therefore probes the extent to which variations in the correlator, $G(\tau;N_\tau)$, are due to the spectral function (i.e. physics) changing rather than the kernel (i.e. geometry).
By studying this ratio, we observe that the positive parity sector has a larger thermal effect than the negative sector \cite{Aarts:2023nax}.

We also define the following integrated ratio to study parity degeneracy \cite{Datta:2012fz},
\begin{equation}
R = \frac{\sum_\tau R(\tau)/\sigma(\tau)^2}{\sum_\tau 1/\sigma(\tau)^2}
\;\;\;\;\;\;
\text{where}
\;\;\;\;\;\;
R(\tau) = \frac{G_+(\tau) - G_-(\tau)}
               {G_+(\tau) + G_-(\tau)},
\label{eq:R}
\end{equation}
where $\sigma(\tau)$ is the variance of $G(\tau)$.
In the degenerate limit, $R\rightarrow 0$, whereas in the limit of extreme non-degeneracy, $R\rightarrow 1$.

\begin{figure}[t!]
  \centering
  \includegraphics[page=1,width=0.48\columnwidth,keepaspectratio,origin=c]
  {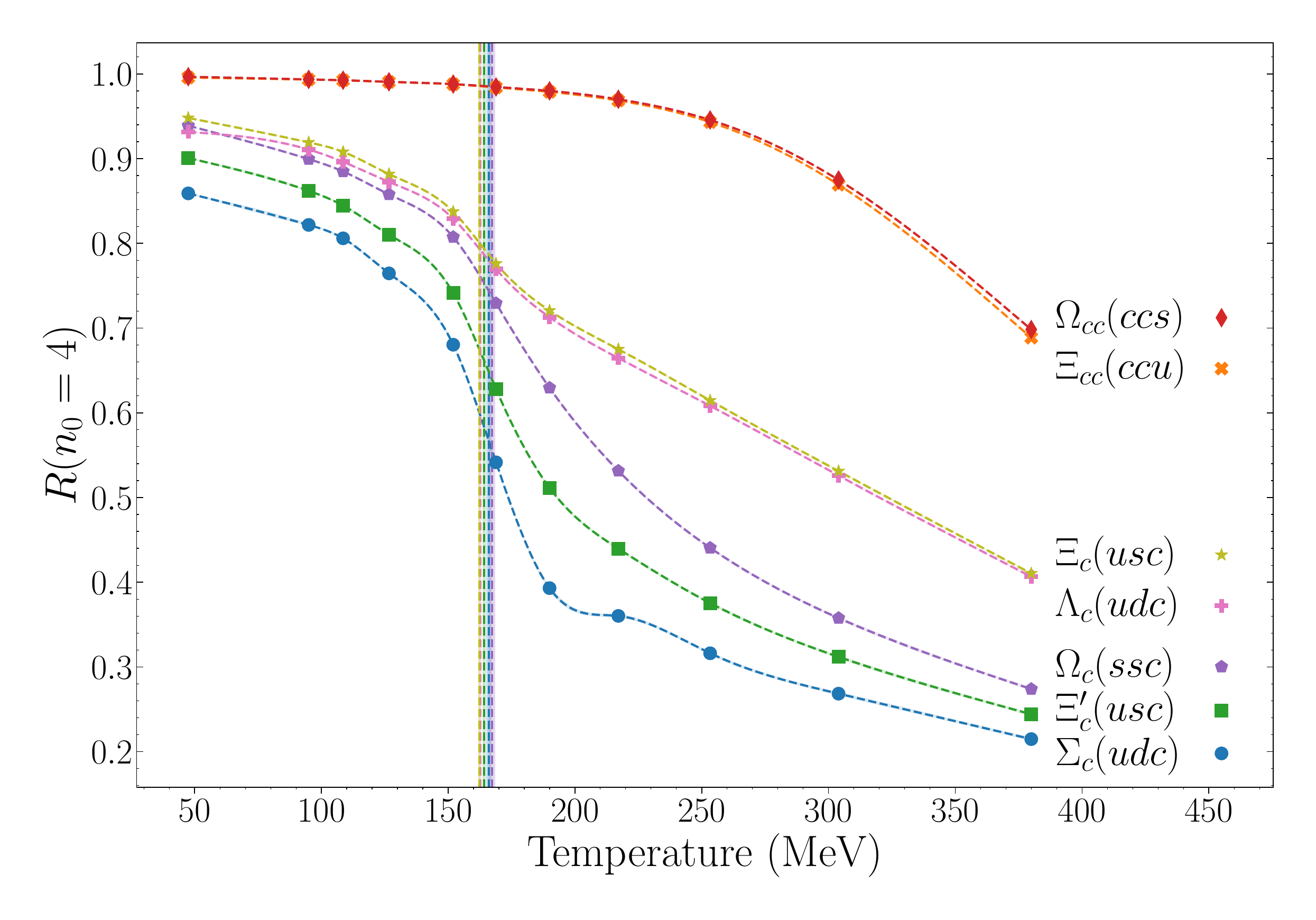}
  \includegraphics[page=1,width=0.48\columnwidth,keepaspectratio,origin=c]
  {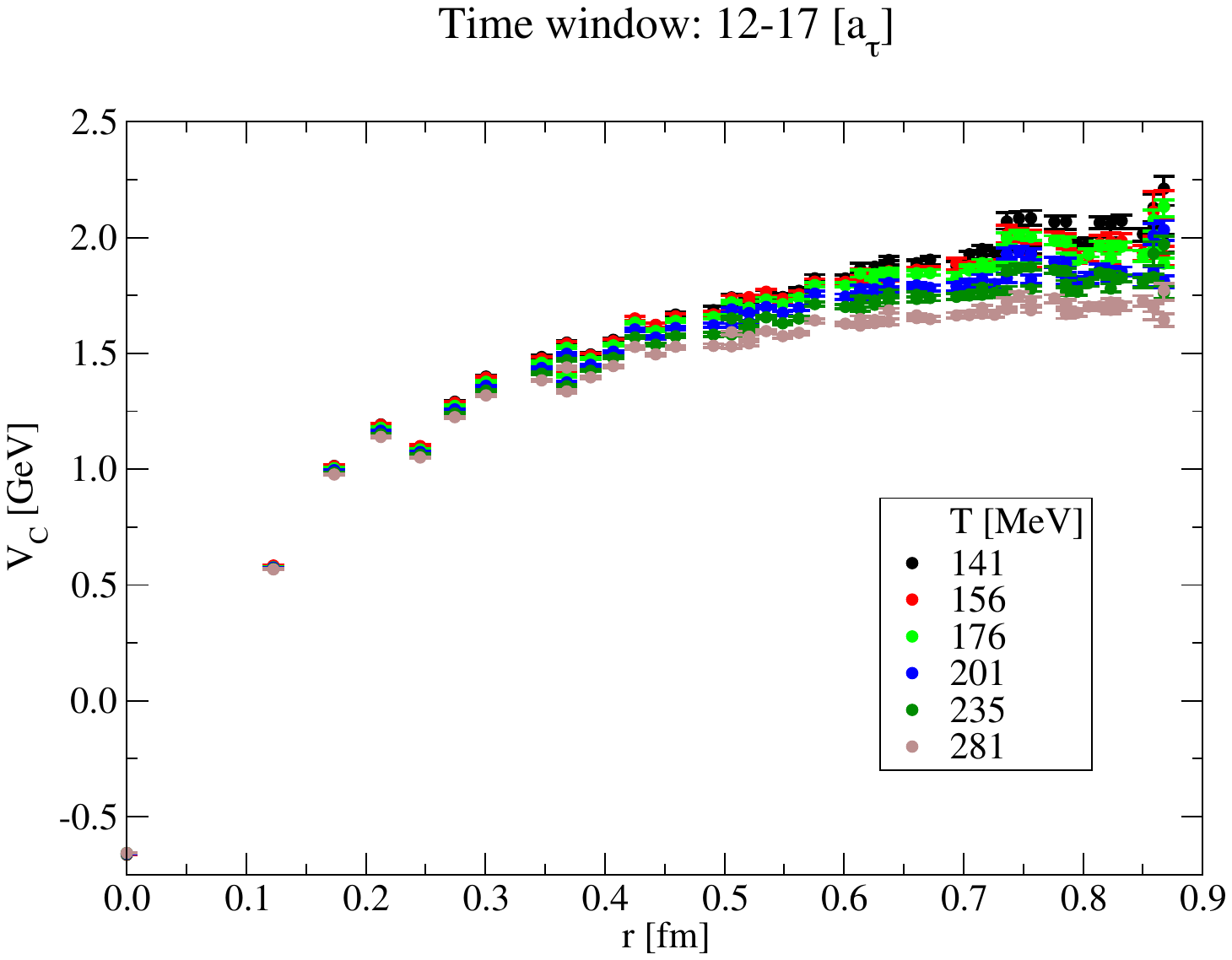}
  \caption{{\bf (Left)} The integrated ratio, $R$ for charm baryons showing the approach towards parity degeneracy as $T$ increases.
  $R$ is defined in Eq.(\ref{eq:R}) and the sum over $\tau$ starts from $n_0$.
  The vertical lines show the points of inflection near $T_{pc}$ for the singly-charmed channels (doubly-charmed channels have no points of inflection).
  \\
  {\bf (Right)} Preliminary results for the interquark potential in (NRQCD) bottomonium mesons for a variety of temperatures above and below $T_{pc}$.
  }
  \label{fig:R-parity-potential}
\end{figure}

In Fig.~\ref{fig:R-parity-potential} (left-panel), we plot $R$ as a function of $T$ for a variety of charmed baryonic channels.
As can be seen, there is clear non-degeneracy for low $T$ and a tendency towards the $R=0$ degenerate limit as $T$ increases or the constituent quark masses decrease.
From the point of inflection, we obtain estimates of the transition temperature. 
Note that this method is applicable only for the singly-charmed channels, because there are no points of inflection for doubly-charmed mesons.
These agree both amongst the channels considered, and with estimates of $T_{pc}$ obtained from the chiral condensate \cite{Aarts:2022krz}.
Thus we have confirmed the picture of approximate parity partner degeneracy above the transition temperature $T_{pc}$.

More extensive discussions of these results, along with parity partner studies of other baryons can be found in 
\cite{
Aarts:2015mma,
Aarts:2017rrl,
Aarts:2018glk,
Aarts:2023nax}.

\section{Interquark Potential in Bottomonium}

One direct way of studying the thermal confinement transition is via the interquark potential.
We apply the {\sc hal-qcd} method \cite{Ishii:2006ec} to obtain the interquark potential in the bottomonium system where the bottom quarks use the NRQCD approximation.
In this approach, the Schr\"odinger equation is ``reverse-engineered'' to derive the potential from the Bethe-Salpeter wavefunctions, which in turn are obtained from non-local mesonic operators \cite{Evans:2013yva}.
A new ``linear regression'' method is used \cite{inpreparation,TomPhD} which
enables more accurate results with better control over systematics.

Figure~\ref{fig:R-parity-potential} (right-panel) has some preliminary results for the potential (using one particular time window for the fits). This shows a clear temperature variation with a reduction in the string tension (i.e. the potential's gradient at large distance) with temperature.

\section{Conclusion}

We present a selection of thermal QCD lattice results from the {\sc fastsum} Collaboration, focusing on recent results for open charm mesons showing clear evidence of thermal effects in the spectrum, charm baryons which uncover a partial restoration of parity degeneracy, and the interquark potential in bottomonium systems where thermal effects in the string tension are seen.

\bigskip
\noindent{\bf Acknowledgements}\\
This work is supported by STFC grant ST/T000648/1 and The Royal Society, with
computing support from DiRAC, EuroHPC, ICHEP, PRACE and the Swansea Academy for Advanced Computing.


\bigskip
\noindent{\bf Authors’ Contributions}\\
Bignell,
García-Mascaraque,
Page,
Spriggs:
Data analysis, plot generation.
J\"ager:
Data production.
Allton,
Aarts,
Anwar,
Burns,
Hands,
Kim,
Lombardo,
Ryan,
Skullerud,
Smecca:
Contributions to analysis and physics interpretations of results.
Allton: draft of the manuscript.


\bibliographystyle{JHEP}
\bibliography{ichep}

\providecommand{\href}[2]{#2}\begingroup\raggedright\begin{thebibliography}{10}

\bibitem{Aarts:2022krz}
G.~Aarts, C.~Allton, R.~Bignell, T.J.~Burns, S.C.~Garc\'\i{}a-Mascaraque et~al., \emph{{Open charm mesons at nonzero temperature: results in the hadronic phase from lattice QCD}},  \href{https://arxiv.org/abs/2209.14681}{{\ttfamily 2209.14681}}.

\bibitem{Aarts:2023nax}
G.~Aarts, C.~Allton, M.N.~Anwar, R.~Bignell, T.J.~Burns et~al., \emph{{Non-zero temperature study of spin 1/2 charmed baryons using lattice gauge theory}}, \href{https://doi.org/10.1140/epja/s10050-024-01261-2}{\emph{Eur. Phys. J. A} {\bfseries 60} (2024) 59} [\href{https://arxiv.org/abs/2308.12207}{{\ttfamily 2308.12207}}].

\bibitem{Aarts:2015mma}
G.~Aarts, C.~Allton, S.~Hands, B.~J\"ager, C.~Praki et~al., \emph{{Nucleons and parity doubling across the deconfinement transition}}, \href{https://doi.org/10.1103/PhysRevD.92.014503}{\emph{Phys. Rev. D} {\bfseries 92} (2015) 014503} [\href{https://arxiv.org/abs/1502.03603}{{\ttfamily 1502.03603}}].

\bibitem{Aarts:2017rrl}
G.~Aarts, C.~Allton, D.~De~Boni, S.~Hands, B.~J\"ager et~al., \emph{{Light baryons below and above the deconfinement transition: medium effects and parity doubling}}, \href{https://doi.org/10.1007/JHEP06(2017)034}{\emph{JHEP} {\bfseries 06} (2017) 034} [\href{https://arxiv.org/abs/1703.09246}{{\ttfamily 1703.09246}}].

\bibitem{Aarts:2018glk}
G.~Aarts, C.~Allton, D.~De~Boni and B.~J\"ager, \emph{{Hyperons in thermal QCD: A lattice view}}, \href{https://doi.org/10.1103/PhysRevD.99.074503}{\emph{Phys. Rev. D} {\bfseries 99} (2019) 074503} [\href{https://arxiv.org/abs/1812.07393}{{\ttfamily 1812.07393}}].

\bibitem{Aarts:2020vyb}
G.~Aarts et~al., \emph{{Properties of the QCD thermal transition with Nf=2+1 flavors of Wilson quark}}, \href{https://doi.org/10.1103/PhysRevD.105.034504}{\emph{Phys. Rev. D} {\bfseries 105} (2022) 034504} [\href{https://arxiv.org/abs/2007.04188}{{\ttfamily 2007.04188}}].

\bibitem{Aarts:2014cda}
G.~Aarts, C.~Allton, T.~Harris, S.~Kim, M.P.~Lombardo et~al., \emph{{The bottomonium spectrum at finite temperature from N$_{f}$ = 2 + 1 lattice QCD}}, \href{https://doi.org/10.1007/JHEP07(2014)097}{\emph{JHEP} {\bfseries 07} (2014) 097} [\href{https://arxiv.org/abs/1402.6210}{{\ttfamily 1402.6210}}].

\bibitem{Aarts:2014nba}
G.~Aarts, C.~Allton, A.~Amato, P.~Giudice, S.~Hands et~al., \emph{{Electrical conductivity and charge diffusion in thermal QCD from the lattice}}, \href{https://doi.org/10.1007/JHEP02(2015)186}{\emph{JHEP} {\bfseries 02} (2015) 186} [\href{https://arxiv.org/abs/1412.6411}{{\ttfamily 1412.6411}}].

\bibitem{Nikolaev:2020vll}
A.~Nikolaev, G.~Aarts, C.~Allton, D.~De~Boni, J.~Glesaaen et~al., \emph{{Mesonic correlators at non-zero baryon chemical potential}}, \href{https://doi.org/10.22323/1.363.0077}{\emph{PoS} {\bfseries LATTICE2019} (2020) 077} [\href{https://arxiv.org/abs/2001.04415}{{\ttfamily 2001.04415}}].

\bibitem{Wilson:2019wfr}
D.J.~Wilson, R.A.~Briceno, J.J.~Dudek, R.G.~Edwards and C.E.~Thomas, \emph{{The quark-mass dependence of elastic $\pi K$ scattering from QCD}}, \href{https://doi.org/10.1103/PhysRevLett.123.042002}{\emph{Phys. Rev. Lett.} {\bfseries 123} (2019) 042002} [\href{https://arxiv.org/abs/1904.03188}{{\ttfamily 1904.03188}}].

\bibitem{Ding:2012sp}
H.T.~Ding, A.~Francis, O.~Kaczmarek, F.~Karsch, H.~Satz et~al., \emph{{Charmonium properties in hot quenched lattice QCD}}, \href{https://doi.org/10.1103/PhysRevD.86.014509}{\emph{Phys. Rev. D} {\bfseries 86} (2012) 014509} [\href{https://arxiv.org/abs/1204.4945}{{\ttfamily 1204.4945}}].

\bibitem{Datta:2012fz}
S.~Datta, S.~Gupta, M.~Padmanath, J.~Maiti and N.~Mathur, \emph{{Nucleons near the QCD deconfinement transition}}, \href{https://doi.org/10.1007/JHEP02(2013)145}{\emph{JHEP} {\bfseries 02} (2013) 145} [\href{https://arxiv.org/abs/1212.2927}{{\ttfamily 1212.2927}}].

\bibitem{Ishii:2006ec}
N.~Ishii, S.~Aoki and T.~Hatsuda, \emph{{The Nuclear Force from Lattice QCD}}, \href{https://doi.org/10.1103/PhysRevLett.99.022001}{\emph{Phys. Rev. Lett.} {\bfseries 99} (2007) 022001} [\href{https://arxiv.org/abs/nucl-th/0611096}{{\ttfamily nucl-th/0611096}}].

\bibitem{Evans:2013yva}
P.W.M.~Evans, C.R.~Allton and J.I.~Skullerud, \emph{{Ab initio calculation of finite-temperature charmonium potentials}}, \href{https://doi.org/10.1103/PhysRevD.89.071502}{\emph{Phys. Rev. D} {\bfseries 89} (2014) 071502} [\href{https://arxiv.org/abs/1303.5331}{{\ttfamily 1303.5331}}].

\bibitem{inpreparation}
T.~Spriggs, C.~Allton, T.J.~Burns and S.~Kim{\emph{,$\,$ in preparation$\!$} }.

\bibitem{TomPhD}
T.~Spriggs{\emph{,$\,$ PhD Thesis, Swansea University$\!\!$} }.

\end{thebibliography}\endgroup

\end{document}